\documentclass[aps,prl,twocolumn,groupedaddress,showpacs]{revtex4}

\usepackage[dvips]{color} 
\usepackage{graphicx}
\usepackage{bm}      
\usepackage{latexsym}
\usepackage{amsmath,amssymb}
\usepackage{color}
\usepackage{stackrel}
\usepackage{accents}

\newcommand{\bsub}{\begin{subequations}}
\newcommand{\esub}{\end{subequations}}

\newcommand{\m}{\bar{M}}
\newcommand{\phih}{\hat{\phi}}
\newcommand{\thetah}{\hat{\theta}}
\newcommand{\Phih}{\hat{\Phi}}
\newcommand{\Thetah}{\hat{\Theta}}
\newcommand{\NO}[1]{:\!\! {#1} \! :}
\newcommand{\Kbar}{\bar{K}}
\newcommand{\rhoh}{\hat{\rho}}
\newcommand{\Jh}{\hat{J}}
\newcommand{\gnd}{{| 0 \rangle}}
\newcommand{\gndmu}{{| 0 \rangle}_{\rho_{0}}}
\newcommand{\gndmuc}{{\phantom{\rangle}}_{\rho_{0}} \langle 0 |}
\newcommand{\dyad}[1]{\accentset{\displaystyle{\leftrightarrow}}{#1}}
\newcommand{\N}{\rho}
\newcommand{\KE}{\mathcal{K}}
\newcommand{\PE}{\mathcal{U}}
\newcommand{\tp}{t'}
\newcommand{\as}{\mathcal{A}_{\sigma}}

\begin{document}

\title{Quantum quench in 1D: Coherent inhomogeneity amplification and `supersolitons'}
\author{Matthew S.\ Foster}
\email{psiborf@rci.rutgers.edu} 
\author{Emil A.\ Yuzbashyan}
\affiliation{Center for Materials Theory, Department of Physics and Astronomy, 
	     Rutgers University, 
	     Piscataway, 
	     NJ 08854, 
	     USA}
\author{Boris L. Altshuler}
\affiliation{Physics Department,
             Columbia University,
             New York,
	     NY 10027,
	     USA}
\date{\today}

\begin{abstract}
We study a quantum quench in a 1D system possessing Luttinger liquid (LL) and 
Mott insulating ground states before and after the quench, respectively. 
We show that the quench induces power law amplification in time of any particle density
inhomogeneity in the initial LL ground state. The scaling exponent is set by the 
fractionalization of the LL quasiparticle number relative to the insulator. As an 
illustration, we consider the traveling density waves launched from an initial localized 
density bump. While these waves exhibit a particular rigid shape, their amplitudes grow 
without bound.
\end{abstract}


\maketitle


The shattering 
of cold glass in hot water is but one of many spectacular
effects that can be induced by a rapid thermal quench in classical
media. What happens when an isolated \emph{quantum} phase of matter
is subject to a sudden, violent deformation of its system Hamiltonian (a 
`quantum quench')? This question is now under vigorous investigation in cold 
atomic gases \cite{Bloch02,Weiss06,Stamper-Kurn06,BlochDalibardZwerger08}.
Long-time, out-of-equilibrium physics already observed in gases confined to 
one \cite{Weiss06}, two \cite{Stamper-Kurn06}, and three \cite{Bloch02} spatial 
dimensions includes oscillatory collapse and revival phenomena \cite{Bloch02,Weiss06} 
and topological defect formation \cite{KibbleZurek,Stamper-Kurn06}.

In this Letter, we study interaction quenches in  one--dimensional (1D) 
quantum
many body systems.
Prior theory assuming spatially uniform dynamics has considered the post-quench distribution of
quasiparticles \cite{Rigol}, correlation functions \cite{CardyCalabrese06,Polkovnikov07}, 
thermalization \cite{Rigol,Rossini09}, quantum critical scaling \cite{Polkovnikov09}, etc. 
On the other hand, the stability of homogeneous solutions with respect to the 
spontaneous eruption of
spatial non-uniformity is by no means  
guaranteed, 
due to the coupling between modes with different
momenta and the extensive 
quantity
of energy 
injected
into the system by the quench. Indeed, homogeneous external 
perturbations are known to generate large spatial modulations in a variety of physical contexts 
\cite{KibbleZurek,Dzero09}.  
We show here that quantum quenches can produce strongly inhomogeneous states
via a mechanism that is ubiquitous in 1D.

\begin{figure}[b]
\includegraphics[width=0.5\textwidth]{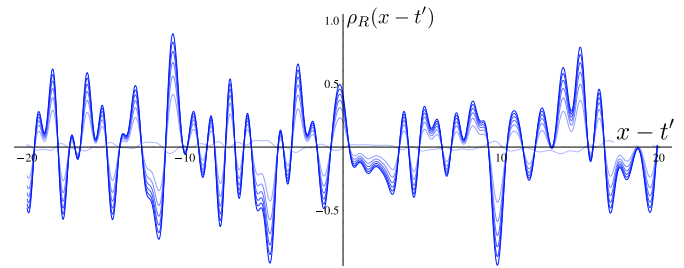}
\caption{
Space-time evolution of the right-moving number density 
$\rho_{R}$
after Luttinger liquid to Mott insulator quench, demonstrating the instability of spatially 
uniform 
dynamics; fainter (bolder) traces depict earlier (later) times.
An infinitesimally small initial density inhomogeneity grows without bound. 
The figure is obtained from Eq.~(\ref{CIA}) 
with $\sigma = 0.8$, $\as = 4.7$, and an initial density profile $\rho_0(x)$ given by a sum of 150 cosines 
with random amplitudes, phases, and wavenumbers.
Amplification occurs for any non-zero $\sigma$, corresponding to a non-zero fractionalization 
of the initial LL quasiparticles with respect to the insulator. 
\label{FigQPF}
}
\end{figure}

\begin{figure}
\includegraphics[width=0.4\textwidth]{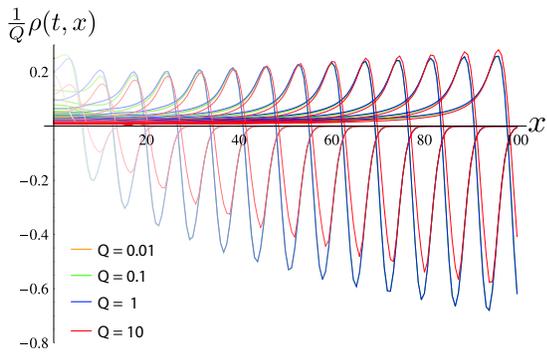}
\caption{
The right-moving  
`supersoliton.' The
number density evolution after Luttinger liquid to Mott insulator quench 
is depicted
as in Fig.~\ref{FigQPF}, 
but here 
for a Gaussian initial 
profile
$\sqrt{\pi}\Delta\rho_{0}(x)=Q\exp{(-x^2/\Delta^2)}$,  
with
$\sigma=0.7$,
now obtained via numerical integration of the exact bosonization result \cite{FYA10}.
Time series for four different $Q$ are plotted; 
the densities are normalized relative to these. 
The evolution is reflection symmetric about $x = 0$. 
\label{FigNumberSuperSol}}
\end{figure}

We consider quenches across a quantum critical point,
with initial (pre-) and final (post-quench) Hamiltonians possessing Luttinger liquid (LL) and Mott insulator ground states, 
respectively. Specifically, we quench into the  insulating phase of the quantum sine Gordon model at the 
``Luther-Emery'' (LE) point \cite{BosonizationRev1,LutherEmery74,Polkovnikov07,Polkovnikov09,IucciCazalilla10}, 
where we are able to determine the dynamics analytically. The pre-quench ground state has an 
inhomogeneous
density 
profile $\rho_0(x)$, which acts as a ``seed'' generating fluctuations in the space-time dynamics of local 
observables
\cite{MosselCaux10,LancasterMitra10}. 
We 
find
that an arbitrarily small deviation of $\rho_0(x)$ 
from a constant is dynamically amplified by the time evolution, see 
e.g.\ Figs.~\ref{FigQPF} and \ref{FigNumberSuperSol}. We argue that the mechanism responsible 
for the amplification is quasiparticle \emph{fractionalization},
a generic attribute
of gapless interacting particles in 1D \cite{BosonizationRev1,FisherGlazman96}. 
We 
further
illustrate the amplification effect
for a 
localized (Gaussian) 
initial density ``bump.'' 
This bump gives rise to a pair of
 non-dispersive, non-interacting density waves that exhibit a rigid shape, 
with amplitudes that grow in time as a power law.
We have dubbed these traveling density waves 
`supersolitons'; an example is depicted in Fig.~\ref{FigNumberSuperSol}.

Specifically, for the Fourier transform $\tilde{\rho}(t,k)$ of the density operator expectation value
$\rho(t,x)$, we find the following exact asymptotic result, valid in the long time limit: 
\begin{equation}\label{CIA}
	\frac{\tilde{\rho}(t,k)}{\tilde{\rho}_{0}(k)}
	=
	\cos(k \tp) 
	-	
	\as \,
	(|k| \tp)^{\sigma/2}
	\cos\left(|k| \tp + \frac{\pi \sigma}{4}\right),
\end{equation}
where $\as$ is a non-universal, 
$k$-independent constant and $\tp \equiv t/\Kbar$, where $\Kbar=1/4$ locates the LE point (see below); 
the quench is performed at $\tp = 0$. The exponent $\sigma$ in 
Eq.~(\ref{CIA}) is determined by the relative fractionalization of the LL quasiparticle 
number with respect to the Mott insulator,
\begin{equation}\label{sigmaDef}
	\sigma \equiv \left( \Kbar/2 K + K/2 \Kbar \right) - 1,
\end{equation}
where $K$ is the Luttinger parameter characterizing the initial Hamiltonian.
Eq.~(\ref{CIA}) implies that the density splits into 
non-dispersing left- and right-moving components, 
$\rho(t,x) = \rho_{R}(x - \tp) + \rho_{L}(x + \tp)$. Interestingly,
the long time response is linear in $\tilde{\rho}_{0}$ and enhanced at shorter wavelengths 
due to the fractional derivative 
($|k|^{\sigma/2}$) 
factor. For $\sigma > 0$, the fluctuations of $\rho_{R,L}$ are continuously amplified by the quench.  
The effect is demonstrated in Fig.~\ref{FigQPF}

In the rest of this Letter, we will explain the setup and calculations 
leading to Eq.~(\ref{CIA}).
Before the quench, our cold atom system is assumed to reside in the ground state 
$\gndmu$ of the LL Hamiltonian
\begin{equation}\label{HInitialDef}
	H_{i} 
	=
	\int dx
	\left[
	\frac{v K}{2} \left(\frac{d \hat{\phi}}{d x}\right)^2
	+
	\frac{v }{2 K} \left(\frac{d \thetah}{d x}\right)^2
	- 
	\frac{\rho_{0}(x)}{q \sqrt{\pi}}\frac{d \thetah}{d x}
	\right],
\end{equation}
where $v$ is the sound velocity, $K$ is the Luttinger parameter, 
and $\rho_{0}(x)/q$ is an external chemical potential, with $q \equiv K /v \pi$.
The 
Hamiltonian in Eq.~(\ref{HInitialDef}) governs the low-energy, long-wavelength 
physics of many gapless 1D cold atomic and condensed matter quantum systems 
\cite{BosonizationRev1,Cazalilla04}; 
in this paper, we have in mind a 1D optical lattice gas of spin-polarized, neutral Fermi atoms, 
but other interpretations are possible. The short-ranged interatomic interactions 
determine $v$ and $K$; repulsive (attractive) interactions correspond to $K < 1$ ($K > 1$), 
while the free Fermi gas has $K = 1$ and $v$ equal to the bare Fermi velocity. The boson fields 
$\phih$ and $\thetah$ encode fluctuations of the long wavelength fermion number density $\NO{\rhoh}$ 
and current $\NO{\Jh}$ on top of the filled Fermi sea via
$\sqrt{\pi} \NO{\rhoh} = d \thetah/d x$ and $\sqrt{\pi} \NO{\Jh} = d \phih/d x$,
where $\NO{\ldots}$ denotes normal-ordering with respect to the \emph{homogeneous}
ground state $\gnd_{\rho_{0} = 0}$. These satisfy the commutation relations 
$[\NO{\rhoh(x)}, \NO{\Jh(x')}] = - (i/\pi) (d/dx) \delta(x - x')$.
The static chemical potential in Eq.~(\ref{HInitialDef}) 
allows us to ``write'' an arbitrary density profile into 
$\gndmu$ via the axial anomaly 
\cite{BosonizationRev1,footnote1},
\begin{equation}\label{AxialAnomaly}
	\gndmuc \NO{\hat{\rho}(x)} \gndmu = \rho_{0}(x),\quad
	\gndmuc \NO{\hat{J}(x)} \gndmu = 0.
\end{equation}


With our system initially prepared in the LL ground state $\gndmu$, 
we perform the quench at time $t = 0$. The dynamics for
$t > 0$ are generated by the translationally invariant, 
``final state'' Hamiltonian $H_{f}$, which favors a gapped, Mott insulating ground state.
Specifically, $H_{f}$ is the Hamiltonian of the quantum sine Gordon model,
\begin{align}\label{HFinalDefB}
	H_{f}
	=
	\frac{1}{K_{f}}
	\int dx \,
	\left[
	\frac{1}{2}  
	\right.&\!
	\left(\frac{d \hat{\Phi}}{d x}\right)^2
	+
	\frac{1}{2} \left(\frac{d \hat{\Theta}}{d x}\right)^2
	\nonumber\\
	&\left.
	\;\;\;
	+	
	\frac{M}{\pi \alpha} \cos\left(2\sqrt{4\pi K_{f}} \hat{\Theta} \right)
	\right].
\end{align}
In Eq.~(\ref{HFinalDefB}) we have expressed $H_{f}$ in terms of the canonically rescaled
boson variables $\Phih \equiv \sqrt{K_{f}} \phih$ and $\Thetah \equiv \thetah/\sqrt{K_{f}}$.
The Mott gap-inducing interparticle interactions set the parameters $M$ and $K_{f}$.
In the context of a Fermi lattice gas at commensurate
filling, the ``Luttinger parameter'' $K_{f}$ characterizes pure forward scattering, 
while $M$ gives the strength of backward scattering Umklapp interactions;
$\alpha$ is a cutoff-dependent 
length scale. 
The ground state of $H_{f}$ is 
gapped for arbitrarily small $M$ over the regime $0 < K_{f} < 1/2$, 
in which the quantum sine Gordon model is integrable \cite{BosonizationRev1}.
The solitons and antisolitons of the classical sine Gordon equation appear as
massive Dirac fermions in the quantum version \cite{Rajaraman}. Solitons repel antisolitons
for $1/4 < K_{f} < 1/2$ and attract them for $0 < K_{f} < 1/4$; in the latter case,
additional bosonic bound states (\emph{breathers}) appear in the spectrum. 
We choose to quench to the boundary between these two regimes,
where $K_{f} = \Kbar \equiv 1/4$. At this special ``Luther-Emery'' point, the 
interactions between the quantum solitons switch off, and $H_{f}$ can be 
\emph{refermionized} \cite{LutherEmery74} in terms of a massive non-interacting 
soliton field $\Psi$,
\begin{align}\label{HFinalDefF}
	H_{f}
	&=
	\frac{1}{\Kbar}
	\int dx \,
	\Psi^\dagger \left(-i \hat{\sigma}^{3} \frac{d}{d x} + \m \hat{\sigma}^2 \right) \Psi.
\end{align}
In this equation, $\Psi$ is a two-component Dirac fermion that is related to
the boson fields in Eq.~(\ref{HFinalDefB}) via the bosonization identity,
$\Psi^{(1,2)} \propto \exp[i \sqrt{\pi} (\Phih \pm \Thetah)]$; 
$\hat{\sigma}^{2,3}$ are Pauli matrices in the standard basis.
The mass gap $\m$ in Eq.~(\ref{HFinalDefF}) is a non-universal, cutoff-dependent quantity. 

It is instructive to rewrite $H_{i}$ [Eq.~(\ref{HInitialDef})] in terms of $\Psi$,
\begin{align}\label{HInitialF}
	H_{i} 
	=
	\int dx \,
	&
	\left\{
	\tilde{v}
	\Psi^\dagger
	\left(
	- i \hat{\sigma}^{3} \frac{d }{d x}
	\right)
	\Psi
	- 
	\frac{\rho_{0}(x)}{2 q} \Psi^\dagger \Psi
	\right.
	\nonumber\\&
	\left.
	\;\;\;
	+
	\frac{\pi \tilde{v}}{2 K^2}
	\left[\Kbar^2 - K^2\right] : \Psi^\dagger \Psi : : \Psi^\dagger \Psi :
	\right\},
\end{align}
where $\tilde{v} \equiv K v/\Kbar$. Comparing Eqs.~(\ref{HFinalDefF}) and (\ref{HInitialF}),
we see that the quench with $K = \bar{K}$ is special.
For this case only (``non-interacting'' quench), the quasiparticles
of the initial and final Hamiltonians are in one-to-one correspondence.
At any other value of $K \neq \bar{K}$ (``interacting quench''), an elementary 
excitation of the initial state carries a fraction of the final state quasiparticle 
number; 
that is, the ``quasiparticle'' excitations of the initial LL 
phase carry $K/\bar{K} = 4 K$ of the global $U(1)$ $\Psi$ fermion number charge
\cite{FisherGlazman96}.
When viewed in terms of $\Psi$, the transition between $H_{i}$ and $H_{f}$ permits a 
dual interpretation as a LL to band insulator quench. 
Correlation functions in the homogeneous quench [$\rho_{0}(x) = 0$] have been previously studied
in Refs.~\cite{Polkovnikov07,IucciCazalilla10}.


To characterize the post-quench dynamics, we compute the expectation values
of the particle number ($\rho$), kinetic ($\KE$) and potential ($\PE$) energy densities 
(the latter two observables are defined with respect to $H_{f}$):
\bsub\label{Observables}
\begin{align}
	\rho(t,x)
	&=
	\frac{1}{2}
	\gndmuc \NO{\Psi^\dagger(t,x) \Psi(t,x)} \gndmu, 
	\label{number}\\
	\KE(t,x)
	&\equiv
	-\frac{i}{2} \gndmuc \Psi^\dagger(t,x) \, \hat{\sigma}^{3} \dyad{\partial_{x}} \, \Psi(t,x) \gndmu, 
	\label{KE}\\
	\PE(t,x)
	&\equiv
	\gndmuc \Psi^\dagger(t,x) \hat{\sigma}^{2} \Psi(t,x) \gndmu,
	\label{PE}
\end{align}
\esub
where $f \dyad{\partial} g \equiv f \partial g - (\partial f) g$. 
In these equations, $\Psi(t,x)$ denotes the Heisenberg picture fermion operator
whose dynamics are generated by $H_{f}$ in Eq.~(\ref{HFinalDefF}).
$\PE$ gives the expectation of the cosine operator in the sine Gordon model 
[Eq.~(\ref{HFinalDefB})], and can be interpreted as a (squared) order parameter for the Mott phase. 
We obtain $\N$, $\KE$, and $\PE$ by solving the Heisenberg equations of
motion for $\Psi(t,x)$ 
and exploiting
the bosonization map. 
Given an arbitrary initial $\rho_{0}(x)$, 
we have derived exact results for 
$\rho$, $\KE$, and $\PE$
at any time $t \geq 0$, 
which will appear elsewhere \cite{FYA10}.


The exact post-quench observables in Eq.~(\ref{Observables}) 
depend upon $\rho_{0}(x)$, $\m$, 
and the dynamic exponent $\sigma$ defined via Eq.~\eqref{sigmaDef}.
 The non-interacting quench with $K = \Kbar$ has $\sigma = 0$, while the
interacting quench ($K \neq \Kbar$) has $\sigma > 0$. We confine ourselves to the range 
$0 \leq \sigma < 1$, for which the $\rho_{0}(x)$-dependent contributions to $\rho$, $\KE$, and 
$\PE$ are given by ultraviolet (UV) convergent integrals \cite{FYA10}. At $\sigma = 1$, these observables 
acquire logarithmic UV divergences, suggesting the onset of a sensitive dependence on lattice 
scale details.
 
We 
now describe
our main results, which concern the $\rho_{0}(x)$-dependent contributions 
to $\rho$, $\KE$, and $\PE$;
the behavior of $\KE$ and $\PE$ for the homogeneous quench $\rho_{0} = \rho(t,x) = 0$ 
will be discussed elsewhere \cite{FYA10}.
The exact leading asymptotic expression for $\rho(t,x)$ in the limit 
$t \rightarrow \infty$ was already given by Eq.~(\ref{CIA}), above. 
Let us specialize this result to a localized initial density profile.
The interacting ($\sigma > 0$) versus non-interacting ($\sigma = 0$) quenches yield
qualitatively different behaviors.
For the interacting quench, Eq.~(\ref{CIA}) implies that $\rho(t,x)$ develops 
a 
non-dispersive response to the 
initial condition for any $0 < \sigma < 1$. 
For example, a Gaussian density bump, $\sqrt{\pi}\Delta\rho_{0}(x)=Q\exp{(-x^2/\Delta^2)}$,   
induces the following asymptotic 
space-time evolution for $t \gg 1/\m$:
\begin{align}\label{SSNumber}
	\rho(t,x)
	=&
	\frac{Q}{2 \sqrt{\pi} \Delta} e^{-\frac{\left( x - \tp \right)^2}{\Delta^2}}
	\nonumber\\
	&
	\begin{aligned}
	-
	&
	\frac{Q}{2 \Delta}	 
	\frac{\Gamma(1 - \sigma)}{\Gamma\left(\frac{1 + \sigma}{2}\right)}
	\left[
	\frac{(K \m \alpha)^2 \, \tp}{\sqrt{2} \Delta}
	\right]^{\sigma/2}
	\\&\times
	e^{-\frac{\left( x - \tp \right)^2}{2 \Delta^2}}
	D_{\sigma/2}\left[\sqrt{2}\left(\frac{x - \tp}{\Delta}\right)\right]
	\end{aligned}
	\nonumber\\
	&+
	\{x \rightarrow -x\},
\end{align}
where $D_{\nu}(x)$ denotes the parabolic cylinder function, $\tp = t/\Kbar$, and we have written out
the explicit form of the prefactor $\as$,
which is non-universal for $\sigma > 0$ and depends upon $\m \alpha$.
The naive continuum calculation gives $\m \alpha = 15/16$.
The divergence of the prefactor at $\sigma = 1$ indicates
the onset of sensitivity to the UV sector of the theory.

Eq.~(\ref{SSNumber}) implies that an antecedent
Gaussian density bump splits into right- and left-moving non-dispersive waves, for
generic $Q$, $\Delta$, and $K \neq \Kbar$ ($\sigma > 0$).
In the long time limit, the leading response is \emph{strictly linear} in $Q$,
with an amplitude that grows as $\tp^{\sigma/2}$. 
Two Gaussian bumps initially separated by a distance $d \gg \Delta$ can be used to create
left- and right-moving waves which pass through each other without changing their form \cite{FYA10}.
We dub these rigid, non-interacting density waves 
`supersolitons'
to distinguish them from the elementary quantum solitons annihilated
by the fermion field $\Psi$. We have confirmed the asymptotic 
result in Eq.~(\ref{SSNumber}) by comparing to numerical integration of the exact
bosonization expression for $\rho$. The supersoliton is 
exhibited in Fig.~\ref{FigNumberSuperSol}.

\begin{figure}[b]
\includegraphics[width=0.4\textwidth]{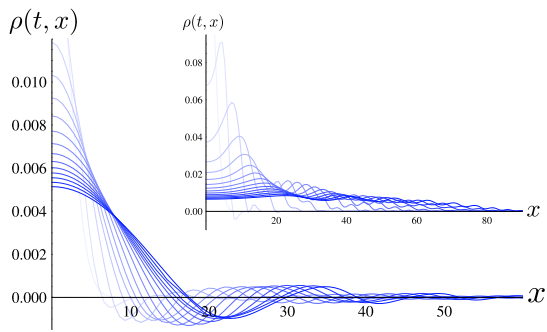}
\caption{
The number density evolution as in Fig.~\ref{FigNumberSuperSol},
but for the non-interacting quench $K = \Kbar$ ($\sigma = 0$). 
The height of the initial bump is $Q = 0.1$ in the main figure and $Q=1$ in the inset;
the evolution is reflection symmetric about $x = 0$. Now there is no fractionalization of the 
initial LL quasiparticles with respect to the insulator and, consequently, the dynamics
are simply dispersive with no supersolitons or inhomogeneity growth.
\label{FigNumberLE}}
\end{figure}

\begin{figure}
\includegraphics[width=0.4\textwidth]{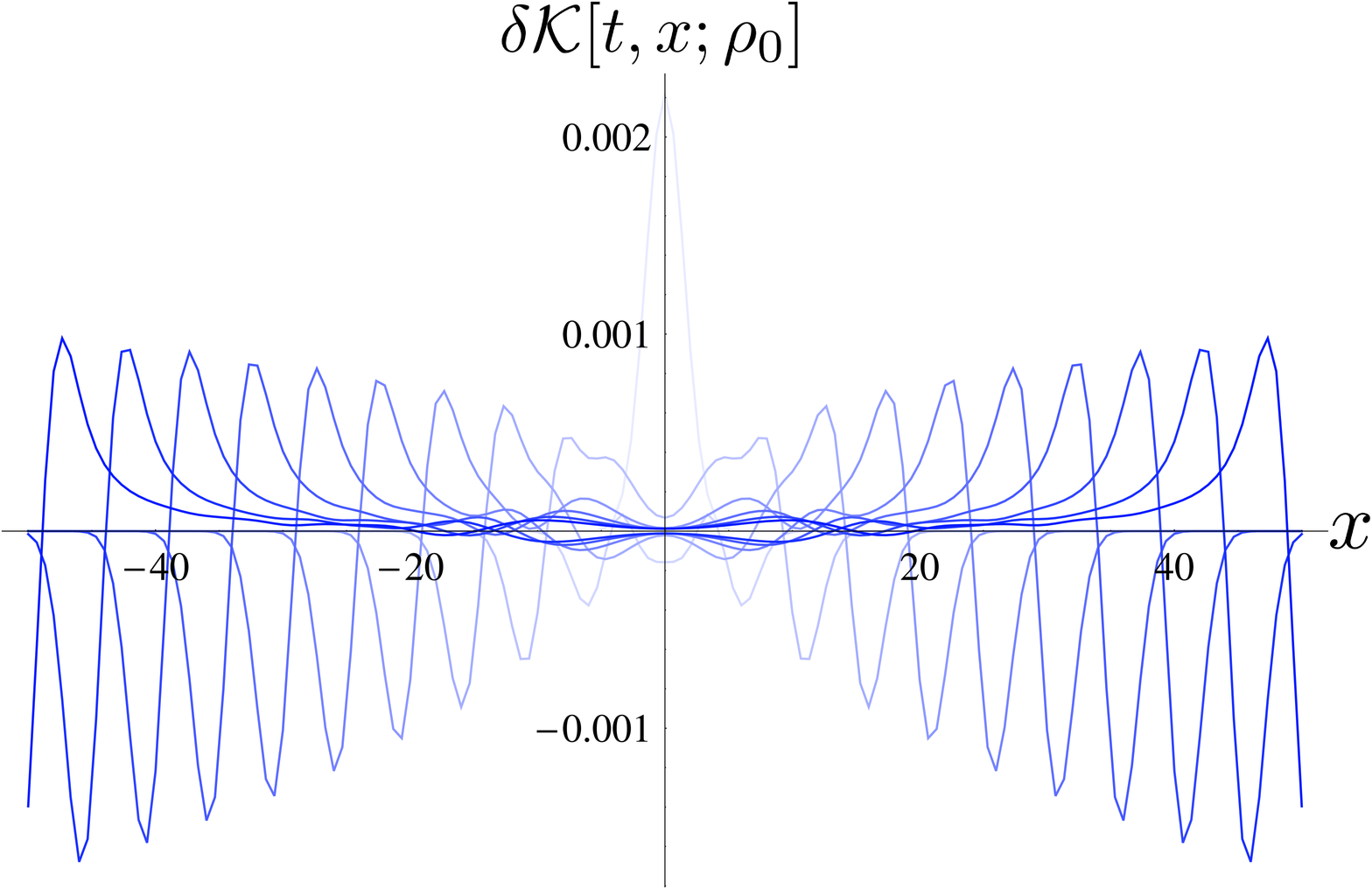}
\caption{The kinetic energy density supersolitons. 
This figure depicts the post-quench space-time evolution of the relative kinetic energy 
density expectation value $\delta\KE[t,x;\rho_0]$ (see text) for $K = 0.77 \neq \Kbar$ and an 
initial Gaussian density bump.
\label{FigKESuperSol}}
\end{figure}

Although the precise shape of the supersoliton implied by Eq.~(\ref{SSNumber}) deforms 
continuously with $\sigma$, it exhibits the same positive-negative ``dipolar'' peak 
profile for any $0 < \sigma < 1$ (see Fig.~\ref{FigNumberSuperSol}). The negative density
dip represents a local evacuation of the filled Fermi sea, which is infinitely deep
in the Luttinger model \cite{BosonizationRev1}. For any $\sigma > 0$, the integral of
the second term in Eq.~(\ref{SSNumber}) over real $x$ vanishes, consistent with 
particle number conservation.
In the limit of the non-interacting quench $\sigma \rightarrow 0$, the right-hand side of
Eq.~(\ref{SSNumber}) vanishes; in this case, the response obtains entirely from subleading terms
that do not grow with $t$ (and conserve the particle number), but which we have not written here.
The same is true in Eq.~(\ref{CIA}) because $\as \rightarrow 1$ when $\sigma \rightarrow 0$.

For comparison, Fig.~\ref{FigNumberLE} depicts the number density $\rho(t,x)$ for the
case $\sigma = 0$, obtained by numerical integration of the 
exact result. 
The main message of this figure is that the non-interacting post-quench dynamics are 
``passive'' and dispersive, depending sensitively upon the details of the initial inhomogeneity 
and showing no amplification phenomena. 

In the interacting quench, the supersoliton is also observed in the relative 
kinetic energy density, defined as $\delta\KE[t,x;\rho_0] \equiv \KE[t,x;\rho_0] - \KE[t,x;0]$,
shown in Fig.~\ref{FigKESuperSol}. 
By contrast, we find that the potential energy density $\PE(t,x)$ does
not exhibit the supersoliton on top of the 
homogeneous
background it acquires after the quench.
The amplification in Eq.~(\ref{CIA}) does not therefore
appear related to a Kibble-Zurek process \cite{KibbleZurek} in the order parameter.

The physical mechanism underlying the power law inhomogeneity growth in Eqs.~(\ref{CIA}) and (\ref{SSNumber})
can be partially elucidated via an analogy to the equilibrium tunneling density of states (TDOS) $\nu(\omega)$ in a LL 
\cite{FisherGlazman96}. Upon tunneling into a one channel quantum wire at $T = 0$ 
characterized by the Luttinger parameter $K$, the conductance at a bias $\omega = e V$ 
diminishes as 
	$
	\nu(\omega) \sim |\omega|^{\sigma}
	$
where $\sigma$ is defined as in Eq.~(\ref{sigmaDef}), but with $\Kbar = 1$. The physics 
behind 
this result
is a follows:
The independent LL ``quasiparticles'' carry a fraction $K$ of the electron charge $e$ 
\cite{FisherGlazman96}.  The TDOS $\nu(\omega)$ vanishes as $\omega \rightarrow 0$ because a 
``whole'' electron must fractionalize into a large number of pieces upon penetrating into the LL, 
and this process is prohibited by phase space restrictions in the low bias limit.
Mathematically, 
the TDOS result
 obtains from the Fourier transform of the electron Green's function in the LL. 
The $t^{\sigma/2}$ amplification in Eq.~(\ref{CIA}) is rendered by a similar mechanism in the quench: 
an initial LL correlation function is convolved with an oscillatory kernel [a product of Green's 
functions resulting from the solution to the Heisenberg equations of motion for $\Psi(t,x)$]. 
The final state Hamiltonian 
$H_{f}$
introduces a scale $\m$, by which the analog of the frequency $\omega$ in the TDOS is the 
evolution interval $\m^2 t$. We might therefore naively expect $\rho(t,x) \sim t^\sigma$, with $\sigma$ defined by 
Eq.~(\ref{sigmaDef}). That the leading power is $\sigma/2$ in Eqs.~(\ref{CIA}) and (\ref{SSNumber})
obtains from a cancelation of $t^\sigma$ terms. 
This suggests that the \emph{immiscibility} of quantum phases composed of quasiparticles carrying relatively 
fractional charges may underlie both the equilibrium TDOS and the quench amplification.

In conclusion, we have shown that a quantum quench can beget a strongly inhomogeneous state,
due to the interplay between quasiparticle fractionalization and the presence
of a mass scale in the final state Hamiltonian. Fractionalization is a robust
feature of 1D gapless phases, so we expect the inhomogeneity proliferation to 
occur in many 1D quantum quenches. 
It would 
be interesting  to consider quenches to final states away from the free fermion LE point
where (super?) soliton-soliton interactions can play a role in the dynamics.

\begin{acknowledgments}

We would like to thank Leon Balents for helpful discussions of LL physics.
This work was 
supported by the National Science Foundation under Award No.~DMR-0547769 and 
by the David and Lucille Packard Foundation.

\end{acknowledgments}



\begin{thebibliography}{99}
\bibitem{footnote1}
	Due to the strictly linear spectrum assumed in the Luttinger model, the scalar 
	potential does not confine or localize the liquid, but merely modulates its density.
\end{thebibliography}
\end{document}